\documentclass[12pt,preprint]{aastex}
\usepackage{multirow}

\shorttitle{FIR and CO in faint z$\sim$6 quasars}
\shortauthors{Wang et al.}

\begin{document}

\title{Far-infrared and Molecular CO Emission From the Host Galaxies of Faint Quasars at z$\sim$6}

\author{Ran Wang\altaffilmark{1,2,11},
Jeff Wagg\altaffilmark{1,3},
Chris L. Carilli\altaffilmark{1},
Roberto. Neri\altaffilmark{4},
Fabian Walter\altaffilmark{5},
Alain Omont\altaffilmark{6},
Dominik. A. Riechers\altaffilmark{7,12},
Frank Bertoldi\altaffilmark{8},
Karl M. Menten\altaffilmark{9},
Pierre Cox\altaffilmark{3},
Michael A. Strauss\altaffilmark{10},
Xiaohui Fan\altaffilmark{2},
Linhua Jiang\altaffilmark{2}}
\altaffiltext{1}{National Radio Astronomy Observatory, PO Box 0, Socorro, NM, USA 87801}
\altaffiltext{2}{Steward Observatory, University of Arizona, 933 N Cherry Ave., Tucson, AZ, 85721, USA}
\altaffiltext{3}{European Southern Observatory, Alonso de C\'ordova 3107, Vitacura, Casilla 19001, Santiago 19, Chile}
\altaffiltext{4}{Institute de Radioastronomie Millimetrique, St. Martin d'Heres, F-38406, France}
\altaffiltext{5}{Max-Planck-Institute for Astronomy, K$\rm \ddot o$nigsstuhl 17, 69117 Heidelberg, Germany}
\altaffiltext{6}{Institut d'Astrophysique de Paris, CNRS and Universite Pierre et Marie Curie, Paris, France}
\altaffiltext{7}{California Institute of Technology, 1200 E. California Blvd., Pasadena, CA, 91125, USA}
\altaffiltext{8}{Argelander-Institut f$\rm \ddot u$r Astronomie, University of Bonn, Auf dem H$\rm \ddot u$gel 71, 53121 Bonn, Germany}
\altaffiltext{9}{Max-Planck-Institut f$\rm \ddot u$r Radioastronomie, Auf dem H$\rm \ddot u$gel 71, 53121 Bonn, Germany}
\altaffiltext{10}{Department of Astrophysical Sciences, Princeton University, Princeton, NJ, USA, 08544}
\altaffiltext{11}{Jansky Fellow}
\altaffiltext{12}{Hubble Fellow}
\begin{abstract}
We present new millimeter and radio observations of nine z$\sim$6 
quasars discovered in deep optical and near-infrared surveys. 
We observed the 250 GHz continuum in eight of the nine objects 
and detected three of them. New 1.4 GHz radio continuum data 
have been obtained for four sources, and one has been detected. 
We searched for molecular CO (6-5) line emission in the three 250 GHz  
detections and detected two of them. Combining with previous millimeter 
and radio observations, we study the FIR and radio emission 
and quasar-host galaxy evolution with a sample of 18 z$\sim$6 
quasars that are faint at UV and optical wavelengths 
(rest-frame 1450 $\rm \AA$ magnitudes of $\rm m_{1450}\geq20.2$). 
The average FIR-to-AGN UV luminosity ratio of this faint 
quasar sample is about two times higher than that of the bright 
quasars at z$\sim$6 ($\rm m_{1450}<20.2$). A fit to the average FIR 
and AGN bolometric luminosities of both the UV/optically 
faint and bright z$\sim$6 quasars, and the average luminosities 
of samples of submillimeter/millimeter-observed quasars at 
z$\sim$2 to 5, yields a relationship of $\rm L_{FIR}\sim {L_{bol}}^{0.62}$. 
Five of the 18 faint z$\sim$6 quasars have been detected at 250 GHz. 
These 250 GHz detections, as well as most of the millimeter-detected optically bright z$\sim$6
quasars, follow a shallower trend of $\rm L_{FIR}\sim {L_{bol}}^{0.45}$ 
defined by the starburst-AGN systems in local and high-z universe. 
The millimeter continuum detections in the five objects 
and molecular CO detections in three of them reveal a few 
$\rm \times10^{8}\,M_{\odot}$ of FIR-emitting warm dust and $\rm 10^{10}\,M_{\odot}$ 
of molecular gas in the quasar host galaxies. 
All these results argue for massive star formation 
in the quasar host galaxies, with estimated star formation rates of a few 
hundred $\rm M_{\odot}\,yr^{-1}$. Additionally, the higher FIR-to-AGN luminosity 
ratio found in these 250 GHz-detected faint quasars also 
suggests a higher ratio between star formation rate and supermassive
black hole accretion rate than the UV/optically most 
luminous quasars at z$\sim$6. 
\end{abstract}
\keywords{galaxies: quasars --- galaxies: high-redshift --- galaxies: starburst --- molecular data 
--- galaxies: active}

\section{Introduction}

Understanding the formation and evolution of the first galaxies and the supermassive 
black holes (SMBHs) they host is a key goal of both observational and theoretical 
astronomy. The first sample of more than twenty bright quasars at z$\sim$6 were selected from 
the imaging data of the Sloan Digital Sky Survey (hereafter SDSS main survey, e.g., Fan et al. 2000, 2006). 
These objects are very bright in quasar UV and optical 
emission (SDSS magnitudes of $\rm 18.74\leq z_{AB}\leq20.42$) and have SMBHs 
with masses of $\rm M_{BH}\sim$ a few $\rm10^{9}\,M_{\odot}$ (Jiang et al. 2007; 
Kurk et al. 2007). 
The dust and molecular gas properties of this bright quasar sample at z$\sim$6 
have been studied at millimeter wavelengths, using the Max Planck Millimeter 
Bolometer Array (MAMBO), the IRAM Plateau de Bure Interferometer (PdBI), 
and the Very Large Array (VLA, Bertoldi et al. 2003a; 2003b; Petric et al. 2003;
Walter et al. 2003, 2004, 2009; Carilli et al. 2004, 2007; Riechers et al. 2009; 
Wang et al. 2007, 2008b, 2010). About 30\% of them have strong continuum 
emission at 250 GHz from 40-to-60 K warm dust (Bertoldi et al. 2003a; Petric et al. 2003;
Wang et al. 2007, 2008b). The FIR luminosities of the MAMBO detections are a
few $\rm \times10^{12}\,L_{\odot}$ to $\rm 10^{13}\,L_{\odot}$, and the 
estimated dust masses are a few $\rm \times10^{8}\,M_{\odot}$ (Beelen et al. 2006; 
Bertoldi et al. 2003a; Wang et al. 2008a). 
Most of the 250 GHz-detected sources have also been detected in molecular
CO line emission with PdBI, yielding molecular gas masses of 
$\rm (0.7-3)\times10^{10}\,M_{\odot}$ (Walter et al. 2003; Carilli et al. 2007; Riechers et al. 2009; 
Wang et al. 2010, 2011). The FIR-CO and FIR-radio luminosity ratios of
these objects are consistent with the trend defined by dusty starburst 
systems at low and high redshifts (Riechers et al. 2006; Beelen et al. 2006; Wang et al. 2008b). 
The results provide direct evidence for massive star
formation at a rate of $\rm \sim10^{3}\,M_{\odot}\,yr^{-1}$,
indicating that bulges are being formed at the same time that the SMBH is rapidly 
accreting in at least 1/3 of the
optically bright quasar systems at z$\sim$6 (Wang et al. 2008b).
Molecular CO (3-2) and (7-6) line emission was resolved in one 
of the FIR and CO luminous quasars, SDSS J114816.64+525150.3 at z=6.42, 
revealing a source size of $\rm \sim5\,kpc$ (Walter et al. 2004, 2009). 
The dynamical mass estimated with the size and line width of the CO 
emission suggest a SMBH-spheroidal bulge mass ratio an order of magnitude 
higher than the typical present-day value (Walter et al. 2004). 

Z$\sim$6 quasars that are fainter at UV and optical wavelengths 
have been discovered from deep optical-near infrared surveys 
(the SDSS southern deep imaging survey, $\rm 20.54\leq z_{AB}\leq22.16$, Jiang et al. 2008, 2009,
the Canada-France High redshift Quasar Survey [CFHQS], $\rm 20.26\leq z_{AB}\leq24.40$, Willott et al. 
2007, 2009a, 2009b). 
Near-infrared spectroscopic observations of some of these  
faint quasars reveal that they are likely to be accreting at the Eddington rate, and have less massive
black holes (i.e., 10$\rm ^{8}$ to $\rm 10^{9}\,M_{\odot}$) than the
SDSS main survey z$\sim$6 quasar sample (Kurk et al. 2009; Willott et al. 2010b).
The steepness of the quasar luminosity function at high redshifts 
is such that lower-luminosity/mass objects are more common. 
Thus, these faint z$\sim$6 quasars provide an important 
sample to understand SMBH-host galaxy formation at the 
earliest epoch. 

The millimeter dust continuum emission from the faint z$\sim$6 
quasars was first studied with a small sample of four objects 
discovered by CFHQS (Willott et al. 2007). One of them was  
detected at 250 GHz with a flux density of 
$\rm \sim1\,mJy$, indicating a FIR continuum luminosity from dust of 
$\rm \gtrsim10^{12}\,L_{\odot}$. The mean 250 GHz 
flux density of the four quasars when stacked is $\rm \sim 0.6\,mJy$, which 
is about a factor of two smaller than the average emission 
of the z$\sim$6 luminous quasar sample (Willott et al. 2007; 
Wang et al. 2008b). We subsequently detected 250 GHz dust continuum and molecular 
CO (6-5) line emission in another two z$\sim$6 quasars with 
$\rm z_{AB}\sim20.7$ (SDSS J205406.42-000514.8 and NDWFS J142516.30+325409.0, 
Jiang et al. 2008; Cool et al. 2006; Wang et al. 2008b, 2010) and 
CO (2-1) line emission from one CFHQS quasar at z=6.2 (CFHQS 
J142952+544717\footnote{We exclude this object from our analysis 
throughout this paper as there is no published millimeter continuum 
observation yet.}, Willott et al. 2010a; Wang et al. 2011). These observations revealed  
a few $\rm 10^{8}\,M_{\odot}$ of 40-to-60 K warm dust and 
$\rm 10^{10}\,M_{\odot}$ of molecular gas in the quasar host galaxies, 
i.e., masses comparable to those found with the bright z$\sim$6 quasars.

In this paper, we report new millimeter and radio observations 
of nine z$\sim$6 quasars. We then study the 
FIR and radio emission with a sample of 18 z$\sim$6 quasars that are 
faint at UV and optical wavelengths, and 
investigate the dust and molecular gas emission properties and star 
forming activity in the host galaxies of the millimeter-detected 
objects. We summarize the current quasar sample at z$\sim$6 and describe 
the new millimeter and radio observations in Section 2, present the 
results in Section 3, analyse the continuum emission properties of the 
faint z$\sim$6 quasar sample in Section 4, discuss the star formation and 
quasar-host evolution in Section 5, and conclude 
in Section 6. A $\rm \Lambda$-CDM cosmology with 
$\rm H_{0}=71km\ s^{-1}\ Mpc^{-1}$, $\rm \Omega_{M}=0.27$ and 
$\rm \Omega_{\Lambda}=0.73$ is adopted throughout this paper 
(Spergel et al. 2007).

\section{Sample and observations}

\subsection{Sample}

There are thirty-three z$\sim$6 quasars that have 
published millimeter and radio observations from the literature (Bertoldi et al. 2003a; 
Petric et al. 2003; Carilli et al. 2004; Wang et al. 2007; 2008b; 
Willott et al. 2007). In this paper, we present new MAMBO 
observations of the 250 GHz dust continuum emission from eight z$\sim$6
quasars (Table 1), and molecular CO observations of three of them.
Seven of the eight sources were discovered from the SDSS 
southern survey with optical magnitudes of $\rm 20.87\leq z_{AB}\leq 22.16$  
(Jiang et al. 2008, 2009), and the other source, ULAS J131911.29+095051.4 
(hereafter J1319+0950), was selected 
from the UKIRT Infrared Deep Sky Survey with $\rm z_{AB}=19.99$ (UKIDSS, Mortlock et al. 2008). 
Seven of the eight objects (all but SDSS J020332.39+001229.3, hereafter J0203+0012) have never been observed at 
millimeter wavelengths before. We also report new radio 1.4 GHz 
continuum observations for three of eight new MAMBO-observed objects and another z$\sim$6 
quasar from CFHQS (Willott et al. 2007). 

The new observations, together 
with the previous data, now provide us a sample of forty 250 
GHz-observed z$\sim$6 quasars with optical magnitudes over 
a wide range, i.e., $\rm 18.74\leq z_{AB}\leq 22.16$, or quasar
rest-frame 1450 $\rm \AA$ AB magnitudes of $\rm 18.81\leq m_{1450}\leq 22.28$. 
Wang et al. (2008b) studied the dust emission from quasars at z$\sim$6, 
mainly based on the most luminous sample from the SDSS main survey. In this paper, we 
investigate the FIR and radio emission properties of the fainter 
quasars at z$\sim$6, i.e., a subsample of 18 objects 
with $\rm m_{1450}\geq20.2$ (Table 1 and 2),  
and compare them to the bright z$\sim$6 quasars ($\rm m_{1450}<20.2$) discussed 
in previous papers (Bertoldi et al. 2003a; Petric et al. 2003; Wang et al. 2008b). 
We will discuss the optical measurements and quasar UV and bolometric luminosities 
based on $\rm m_{1450}$, 
which is derived from the SDSS z-band magnitudes and deep optical spectra (e.g., 
Fan et al. 2000, 2001, 2003, 2004, 2006; Jiang et al. 2008, 2009; Willott et al. 2007), 
or from infrared photometry in the cases in which significant dust reddening is presented in the optical 
spectra (e.g., McGreer et al. 2006).

\subsection{Observations}

Observations of the 250 GHz dust continuum from the eight z$\sim$6 quasars 
were carried out using the MAMBO-II 117-element array  
on the IRAM 30-m telescope (Kreysa et al. 1998) in the winters of 2008-2009 and 2009-2010. We adopted 
the standard on-off photometry mode with a chopping rate of 2 Hz by $\rm 32''$ in 
azimuth (see also Wang et al. 2007, 2008). 
The on-source observing time was about 1-3 hours for each 
object, to achieve 1$\rm\sigma$ rms sensitivities of $\rm \sim0.5$ 
to 0.7 mJy. 
We used the MOPSIC pipeline (Zylka 1998) to reduce the data, 
and three sources, J1319+0950, 
SDSS J012958.51$-$003539.7 (hereafter J0129$-$0035), and 
J0203+0012, were detected at $\rm \ge4\sigma$. Wang et al. (2008b) observed  
J0203+0012 with MAMBO and did not detect the source with a 1$\rm\sigma$ rms of 0.7 mJy.  
Our new observations, together with the old data, improve the rms 
to 0.46 mJy for this object, and we detect the 250 GHz continuum at 4$\rm\sigma$. 
We have also obtained a 2.4$\sigma$ signal from another 
source, SDSS J023930.24$-$004505.4 (hereafter J0239$-$0045).  

We also observed four z$\sim$6 quasars at 1.4 GHz in Winter 2008-2009, 
using the VLA in A configuration with a synthesized beam size  
of $\rm FWHM\sim1.4''$. Flux scales were measured using the 
standard VLA calibrators of 3C286 and 3C48, and phase calibrations were performed 
every 30 min on nearby phase calibrators (1254-116, 1507-168, 2136+006, and 2323-032). 
The typical rms noise level 
was $\rm \sim17\,\mu Jy\,beam^{-1}$ in three hours of observing time 
for each source. We reduced the data with the standard
VLA wide field data reduction software AIPS, and one source, J1319+0950, 
was detected at $\rm \sim3.8\sigma$. 

We searched for molecular CO (6-5) line emission from the 
three MAMBO detections using the 3 mm receiver on 
the PdBI. The source J1319+0950 was observed in Summer 2009 in C and D 
configurations (i.e. a spatial resolution of $\rm FWHM\sim3.5''$) 
with the narrow-band correlator, which provides a bandwidth of 1 GHz 
in dual polarization. 
The observations of J0129$-$0035 and J0203+0012 were
carried out using the new wide band correlator WideX with a
bandwidth of 3.6 GHz in dual polarization, which covers well the 
$\rm Ly\alpha$-redshift uncertainty of $\rm \Delta z\sim\pm0.03$ (Jiang et al. 2009). 
The observations were performed in Summer 2010 in D configuration with a
synthesized beam size of $\rm FWHM\sim5''$. Phase calibration 
was performed every 20 min with observations of phase calibrators, 1307+121 and 
0106+013, and we observed MWC 349 as the flux calibrator. 
We reduced the data with the
IRAM GILDAS software package (Guilloteau \& Lucas 2000).
The CO (6-5) line emission is detected from J1319+0950 at 4.8$\sigma$ 
in 10.3 hours of on-source time with an rms sensitivity 
of $\rm 0.50\,mJy\,beam^{-1}$ per $\rm 70\,km\,s^{-1}$ 
binned channel, and from J0129$-$0035 at 5.3$\sigma$ in 8.8 
hours with an rms of $\rm 0.55\,mJy\,beam^{-1}$ per $\rm 100\,km\,s^{-1}$ 
channel. We also detect the continuum at $\rm>3\sigma$ at the CO (6-5) line 
frequency in both objects. The other source, J0203+0012, was  
undetected in both CO and 3 mm continuum in 13.9 hours of on-source 
time, with a sensitivity of $\rm 0.44\,mJy\,beam^{-1}$ 
per $\rm 100\,km\,s^{-1}$ channel. 

\section{Results}  

\subsection{Notes for individual objects}
We summarize the new millimeter and radio observations of the 
nine z$\sim$6 quasars in Table 1. Detailed measurements of the three new 250 GHz dust 
continuum/molecular CO (6-5) detections are listed below: 

{\bf J1319+0950} The source was discovered in the UKIRT Infrared 
Deep Sky Survey (UKIDSS) with $\rm m_{1450}=19.65$, 
which lies in the typical magnitude 
range of the optically bright z$\sim$6 quasars selected from the SDSS 
main survey (Mortlock et al. 2008). We detect the 250 GHz dust  
continuum emission from this object with a flux density 
of $\rm S_{250 GHz}=4.20\pm0.65\,mJy$, which is the third-strongest 250 GHz 
detection among the known z$\sim$6 quasars. The 1.4 GHz radio continuum 
is also detected from our VLA observation, with $\rm S_{1.4 GHz}=64\pm17\,\mu Jy$ (left panel of Figure 1). 
The radio emission appears unresolved with a peak at $\rm 13^h19^m11.30^s$, 
$\rm 09^{\circ}50'51.8''$, i.e. within $0.4''$ of the optical quasar position.
We also see a 4$\sigma$ peak ($\rm S_{1.4 GHz}=68\pm17\,\mu Jy$) 
at $\rm 13^h19^m11.65^s$, $\rm 09^{\circ}50'53.2''$ 
on the radio map, which is $\rm 5.6''$ away from the quasar. 
For comparison, only 0.02 detections with 
$\rm S_{1.4 GHz}\geq60\,\mu Jy$ are expected from a random circular 
sky area with a radius of $\rm 6''$, according to the 1.4 GHz radio 
source counts from previous deep VLA observations (i.e., 
$\rm N(>S_{1.4 GHz})=(0.40\pm0.04)(S_{1.4 GHz}/75\,\mu Jy)^{(-1.43\pm0.13)}$, Fomalont et al. 2006).
No optical counterpart has been found at the radio source position in the NASA/IPAC 
Extragalactic Database or the SDSS images. 

We observed the CO (6-5) line emission from this object with 
the PdBI. The line is detected at 4.8$\sigma$ on the velocity-integrated 
map with a line flux of $\rm 0.43\pm0.09\,Jy\,km\,s^{-1}$, $\rm 0.9''$ 
away from (i.e., consistent with) the quasar position (the upper panel of Figure 2). 
A Gaussian fit to the line spectrum yields a 
peak line flux density of $\rm 0.80\pm0.15\,mJy\,beam^{-1}$ and a 
line width of $\rm FWHM=537\pm123\,km\,s^{-1}$, centered at the redshift 
of $\rm z=6.1321\pm0.0012$. This is slightly higher than, but consistent 
with the Mg II redshift of $\rm z=6.127\pm0.007$, 
considering the measurement errors ($\rm \pm0.004$, Mortlock et al. 2008) and the typical velocity shift and scatters 
$\rm 97\pm269\,km\,s^{-1}$ ($\rm -0.002\pm0.006$ in redshift) 
between the broad Mg II line emission and the quasar systemic redshift (Richards et al. 2002).
We also detect the continuum under the CO (6-5) line emission 
with a peak value of $\rm S_{97 GHz}=0.31\pm0.08\,mJy\,beam^{-1}$ on the continuum 
map averaged over the line-free channels (right panel of Figure 1). We adopt this as the continuum flux 
density as the source is unresolved with the $\rm 5''$ beam. 
We checked the position of the 1.4 GHz radio source $\rm 5.6''$ away 
from the quasar in the PdBI CO/97 GHz continuum images. There is no 
clear ($\rm \geq3\sigma$) detection at this position and the 
measurement in the 97 GHz continuum image  
is $\rm 0.19\pm0.08\,mJy$ (right panel of Figure 1).

{\bf J0129$-$0035} This is the faintest of the 
SDSS z$\sim$6 quasars, with $\rm m_{1450}=22.16$ (Jiang et al. 2009). 
We detect 250 GHz dust continuum from 
the quasar host galaxy, with a flux density 
of $\rm S_{250 GHz}=2.37\pm0.49\,mJy$, comparable to 
the flux densities of the optically 
bright z$\sim$6 quasars with 250 GHz detections. 

We detected the CO (6-5) line in this source with a flux 
of $\rm 0.37\pm0.07\,Jy\,km\,s^{-1}$, i.e., at 5.3$\sigma$, integrated over a velocity range 
of 380 $\rm km\,s^{-1}$ (the middle panel of Figure 2). The line width fitted with a single 
Gaussian is $\rm 283\pm87\,km\,s^{-1}$, with a line peak of 1.2$\pm$0.3 mJy. 
The host galaxy redshift measured with the CO line is $\rm z=5.7794\pm0.0008$, 
which is in good agreement with the optical redshift of $\rm z_{Ly\alpha}=5.78\pm0.03$ 
measured with $\rm Ly\alpha$ (Jiang et al. 2009). 
The underlying dust continuum is also detected with a flux 
density of $\rm S_{102 GHz}=0.14\pm0.04\,mJy$. 

{\bf J0203+0012} This object is a broad absorption line (BAL) quasar 
discovered by both the SDSS southern survey (Jiang et al. 2008) and 
the UKIDSS survey (Venemans et al. 2007; Mortlock et al. 2008) with $\rm m_{1450}=20.97$. 
The redshift measured with the quasar $\rm N\,\small V$ line emission is z=5.72 (Mortlock et al. 2008). 
Wang et al. (2008b) detected the 1.4 GHz radio continuum  
with $\rm S_{1.4 GHz}=195\pm22\,\mu Jy$ (Wang et al. 2008b).
Our MAMBO observation yields a 250 GHz dust continuum flux density of 
$\rm S_{250GHz}=1.85\pm0.46\,mJy$. We have searched for the CO (6-5) line 
emission from this object over a wide redshift range, from z=5.60 to 5.84, 
but did not detect it in 14 hours of on-source time.  
Assuming a line width of 600 $\rm km\,s^{-1}$, the 3$\rm \sigma$ upper limit of the 
CO flux is estimated to be $\rm 0.32\,Jy\,km\,s^{-1}$ (the lower panel of 
Figure 2)\footnote{We estimate the 3$\sigma$ upper limit for the CO line flux 
as $\rm 3\sigma_{channel}(\Delta V_{line}\Delta V_{channel})^{1/2}$, 
where $\rm \Delta V_{line}$ is the assumed line width, and $\rm \sigma_{channel}$ 
and $\rm \Delta V_{channel}$ are the rms per channel and channel width listed in Section 2.2 
(Seaquist et al. 1995).}. The dust 
continuum at the CO (6-5) line frequency is also undetected with a continuum sensitivity 
of $\rm 0.04\,mJy\,beam^{-1}$ and a 3$\sigma$ upper limit of 0.12 mJy.

\subsection{FIR and radio luminosities and dust mass}

We calculate the FIR luminosities and upper limits for the eight z$\sim$6
quasars observed by MAMBO. We model the FIR emission with an optically 
thin graybody and normalize the model to the MAMBO-250 GHz
and available PdBI-3mm dust continuum flux densities/3$\sigma$ 
upper limits (see also Wang et al. 2007, 2008). 
A dust temperature of $\rm T_d=47\,K$ and emissivity index 
of $\rm \beta=1.6$ are adopted in the model; these are typical values 
for the FIR and CO luminous quasars at lower redshifts (Beelen et al. 2006).
We then integrate the model FIR SED from $\rm 42.5\,\mu m$ to $\rm 122.5\,\mu m$ to 
calculate the total FIR luminosities/upper limits. The FIR luminosity of 
the strongest MAMBO detection, J1319+0950, 
is $\rm 1.0\times10^{13}\,L_{\odot}$, making it one of the most 
FIR-luminous objects among the z$\sim$6 quasar sample. 
The other two MAMBO-detected objects are from the SDSS 
southern survey sample and have FIR luminosities 
of $\rm 4\sim5\times10^{12}\,L_{\odot}$. 
We also estimate the dust masses for the three MAMBO
detections, using $\rm M_{dust}=L_{FIR}/4\pi\int\kappa_{\nu}B_{\nu}d\nu$,
where $\rm B_{\nu}$ is the Planck function and
$\rm \kappa_{\nu}=\kappa_{0}(\nu/\nu_{0})^{\beta}$ is the dust
absorption coefficient with $\rm \kappa_{0}=18.75\,cm^2g^{-1}$
at 125$\,\mu$m (Hildebrand 1983). The estimated dust masses of the three sources
are $\rm (2.5-5.7)\times10^{8}\,M_{\odot}$. 
We finally calculate the 1.4 GHz radio continuum luminosity densities and 
upper limits for the four VLA observed objects. We assume a power-law continuum 
of $\rm f_{\nu}\sim\nu^{-0.75}$ (Condon 1992) to correct the observed frequency to the quasar 
rest frame. The results are summarized in Table 3; we also include the 1.4 GHz radio luminosity 
of J0203+0012 from Wang et al. (2008b). 

\subsection{CO luminosities and molecular gas masses}

We calculate the CO (6-5) line luminosities and upper limits for
the three CO-observed z$\sim$6 quasars, using
$\rm L'_{CO(6-5)}=3.25\times10^{7}I\Delta v_{CO(6-5)}{\nu_{obs}}^{-2}{D_{L}}^2(1+z)^{-3}$ 
($\rm K\,km\,s^{-1}\,pc^2$),
where $\rm I\Delta v_{CO(6-5)}$ is the line flux or 3$\sigma$ upper
limit in $\rm Jy\,km\,s^{-1}$ presented in Table 1, $\rm \nu_{obs}$
is the observed frequency of the CO (6-5) line in GHz, $\rm D_L$
is the luminosity distance in Mpc, and z is the quasar redshift (Solomon \& Vanden Bout 2005).
We then assume a CO excitation ladder similar to that of
J1148+5251 (Riechers et al. 2009), for which 
$\rm L'_{CO(6-5)}/L'_{CO(1-0)}\approx0.78$, allowing us to calculate the CO (1-0) luminosities.
The FIR-to-CO(1-0) luminosity 
ratios of J1319+0950 and J0129$-$0035 are 520 and
340 $\rm L_{\odot}/K\,km\,s^{-1}\,pc^{2}$, respectively, which are 
similar to the values found for other z$\sim$6 CO-detected quasars (Wang et al. 2010). 
We then calculate the molecular gas masses ($\rm M_{gas}=M[H_2+He]$), adopting a CO luminosity-to-gas mass 
conversion factor of $\rm \alpha=0.8 M_{\odot}\,(K\,km\,s^{-1}\,pc^{2})^{-1}$ seen 
in low-redshift ultra-luminous infrared galaxies (ULIRGs, Solomon et al. 1997;
Downes \& Solomon 1998). The molecular gas masses for the two new CO detections are all of order 
$\rm 10^{10}\,M_{\odot}$ (Table 3). 

\section{Analysis} 

The new observations increase the sample of 
MAMBO-observed z$\sim$6 quasars to forty objects (Bertoldi et al. 2003a; Petric et al. 2003;
Willott et al. 2007; Wang et al. 2007, 2008b). 
In this paper, we study their FIR and radio emission properties, focusing on 
the 18 sources that are fainter at UV and optical wavelengths 
($\rm 20.20<m_{1450}\leq22.28$, hereafter faint z$\sim$6 quasars) and compare them with  
the first sample of millimeter-studied z$\sim$6 quasars from 
the SDSS main survey (hereafter bright z$\sim$6 quasars, Bertoldi et al. 2003a; 
Petric et al. 2003; Wang et al. 2007, 2008b). 
The MAMBO observations of these faint z$\sim$6 quasars have typical 
sensitivities (0.4 to 0.6 mJy) comparable to or slightly lower 
than that achieved in previous MAMBO observations of the SDSS 
main survey z$\sim$6 quasars ($\rm 0.5-1.1\,mJy$, Petric et al. 2003;
Wang et al. 2007, 2008) and the optically bright quasar samples at
redshifts 2 and 4 (Omont et al. 2001,2003; Carilli et al. 2001).
Five of the 18 sources are detected at 250 GHz at $\rm \geq3\sigma$ (Wang et al. 2007, 2008b; 
Willott et al. 2007), yielding a detection rate of 28$\pm$12\%, 
or $\rm 25\pm14$\% if we consider the subsample of twelve objects from the SDSS southern 
survey selected with the same color selection criteria (three detections, Fan et al. 2004; 
Jiang et al. 2008; 2009; Wang et al. 2008b, this work). 
These are slightly lower than the MAMBO detection 
rate of $\rm \gtrsim$30\% found with the optically bright quasar 
samples from z=2 to 6 (Omont et al. 2001, 2003; Carilli et al. 2001;
Wang et al. 2008b), but are consistent within the errors.

We searched for molecular CO (6-5) line emission in four of the five 
MAMBO-detected faint z$\sim$6 quasars, and detected three 
(Wang et al. 2010), indicating that highly excited molecular
gas is ubiquitous in these millimeter (FIR) luminous quasar-galaxy systems 
at the earliest epoch. Fourteen of the 18 sources have radio 1.4 GHz 
continuum observations with typical 1$\sigma$ rms of 
$\rm \sim20\,\mu Jy$, and only two of them, J0203+0012 and J1427+3312\footnote{Note 
that J1427+3312 was selected from radio data (McGreer et al. 2006).} 
have radio detections (McGreer et al. 2006; 
Wang et al. 2007, 2008b, this work). This gives a much lower radio 
detection rate for the faint z$\sim$6 quasar sample than that of the 
bright objects; nine out of 22 z$\sim$6 quasars with $\rm m_{1450}<20.2$ were detected 
at 1.4 GHz with similar observational sensitivities ($\rm \sim20\,\mu Jy$, Petric et al. 2003; Carilli et al. 2004; 
Wang et al. 2007, 2008b). The 1.4 GHz continuum flux density is 1.7 mJy for 
J1427+3312 (McGreer et al. 2006) and 195 $\rm \mu$Jy for J0203+0012 (Wang et al. 
2008b), indicating radio-loud nuclei in the two objects\footnote{The radio-loud quasars
are defined as quasars with radio-to-optical flux density ratios
of $\rm f_{\nu,5GHz}/f_{\nu,4400\AA}\geq10$ (Kellermann et al. 1989)  
where fluxes are measured in the rest frame.
We adopt this definition and calculate R assuming a power-law
UV-to-optical continuum of $\rm f_{\nu}\sim\nu^{-0.5}$
(Fan et al. 2000) and radio continuum of $\rm f_{\nu}\sim\nu^{-0.75}$
(Condon 1992). } (Momjian et al. 2008). Deep observations are required to better constrain 
the radio emission from the radio-quiet UV/optical faint z$\sim$6 quasars.

\subsection{The average FIR and radio emission}

We calculate the average 250 GHz and 1.4 GHz continuum for the 18 
faint z$\sim$6 quasars. The mean flux densities are derived for i) 
all sources, ii) MAMBO detections, and iii) MAMBO non-detections, 
using the MAMBO and VLA measurements weighted by 
inverse variance (See also Wang et al. 2008b). We also present the average 
values for the 22 bright quasars ($\rm m_{1450}<20.2$) for comparison. The average emission 
of this bright quasar sample is consistent with the
results presented in Wang et al. (2008b). We list all the results
in Table 4.

The mean 250 GHz flux density 
is $\rm 0.85\pm0.12\,mJy$ for all 18 faint quasars, and $\rm 1.85\pm0.20\,mJy$ for
the five MAMBO-detected sources, which are about 50\% and 70\% 
of the values found for the bright quasars, respectively. The results are 
about the same when the two radio-loud quasars 
are excluded. The stacking measurement for the 13 MAMBO-undetected
sources in the faint sample is $\rm 0.33\pm0.15\,mJy$, which places an upper
limit of $\rm <0.78\,mJy$ (stacking average plus the 3$\rm \sigma$ rms) 
for the average millimeter dust continuum emission. 

We convert the average 250 GHz flux densities to $\rm 42.5--122.5\,\mu m$ FIR luminosities, adopting
the assumptions described in Section 3.2, and calculate the 
average FIR-to-1450$\AA$ luminosity ratios 
($\rm \left\langle L_{FIR}\right\rangle/\left\langle L_{1450}\right\rangle$) 
for each sample (see Table 4).
The average luminosity ratios for all the sources and the MAMBO detections   
in the faint-quasar sample are two and three times higher than the 
corresponding values found with bright quasars. This is consistent 
with the non-linear FIR-to-AGN luminosity relationships of 
$\rm L_{FIR}\sim L_{AGN}^{\alpha}$, where $\rm \alpha<1$ found with 
the AGN-starburst at low and high redshifts (Beelen et al. 2004; 
Hao et al. 2005, 2008; Wang et al. 2008b; Serjeant \&
Hatziminaoglou 2009; Lutz et al. 2010), as we will discuss further 
in Section 4.2. However, the different average luminosity ratios found with  
the MAMBO detections in the bright and faint quasar samples may be 
partly due to the selection effect that the detection limit   
of the MAMBO observations are about the same for the two samples, while 
the AGN luminosities are different by design. 

The average 1.4 GHz continuum flux density, derived 
from 14 VLA-observed sources in the faint quasar sample, is 
$\rm 19\pm6\,\mu Jy$, and $\rm 2\pm6\,\mu Jy$ when the two radio-loud sources
are excluded. This suggests an upper limit of $\rm 20\,\mu Jy$ for 
the radio-quiet and UV/optically faint z$\sim$6 quasars. 
We estimate the average rest-frame 1.4 GHz luminosities/upper limits 
for the two samples and list the results in Table 4. 
We plot $\rm L_{FIR}$ vs. $\rm L_{\nu,1.4GHz}$ of all the MAMBO-detected
z$\sim$6 quasars together with the average luminosities of the
MAMBO-detections and non-detections in Figure 3, and compare them
to sample of IRAS-selected local star forming galaxies (Yun et al. 2001).
The local star forming galaxies with FIR and radio emission powered
by the active star formation have an average luminosity
ratio of $\rm q=log\left(L_{FIR}/3.75\times10^{12}\,L_{\odot}\right)-
log\left(L_{1.4GHz}/L_{\odot}\cdot Hz^{-1}\right)=2.34$. The range
$\rm 1.6<q<3.0$ contains 98\% of the IRAS galaxies (Yun et al. 2001). 
The average luminosities/upper limits of the MAMBO-detected z$\sim$6 
quasars, though lower than the
mean value of $\rm q=2.34$, are all within this range defined by local 
star forming galaxies, This argues that, in addition to the AGN
power, star formation also contributes significantly to the FIR-to-radio emission of
the MAMBO-detected z$\sim$6 quasars. The average ratio of the MAMBO-undetected bright z$\sim$6 
quasars is offset from this range. 

\subsection{The FIR-to-AGN luminosity relationship}

The correlations between FIR luminosities and AGN UV, optical, and bolometric 
luminosities have been studied widely with samples of infrared and optically 
selected quasars to understand the origin of quasar FIR emission. 
Study of samples of the Palomar-Green (PG) quasars and Seyfert galaxies shows a 
relationship of $\rm L_{FIR}\sim L_{AGN}^{0.7\sim0.8}$ for typical 
optically-selected AGN/quasars in the local universe (Hao et al. 2005;
Netzer et al. 2007; Lutz et al. 2010). A shallower relationship 
of $\rm L_{FIR}\sim L_{AGN}^{0.4}$ was found for the 
local IR-luminous quasars hosted by starburst ULIRGs 
(Hao et al. 2005) and high-redshift FIR-and-CO luminous 
quasars (Beelen et al. 2004; Hao et al. 2008; Wang et al. 2007; Lutz et al. 2010), 
which suggests excess FIR emission powered by extreme starburst in these 
systems. Wang et al. (2008b) reported that the millimeter-detected 
bright z$\sim$6 quasars follow the shallow luminosity trend defined 
by the local IR quasars. 

In this work, we investigate the FIR-to-AGN luminosity correlation 
of the full sample z$\sim$6 quasars, and compare with 
the samples of MAMBO-250 GHz or SCUBA-350
GHz observed optically bright quasars at z$\sim$2 to 5 
(Omont et al. 2001, 2003; Carilli et al. 2001;
Priddey et al. 2003a), and local IR quasars 
(Zheng et al. 2002; Hao et al. 2005). We convert the UV 
or optical luminosities of all objects to AGN bolometric luminosities 
assuming a UV-to-optical power-law continuum of $\rm f_{\nu}\sim\nu^{-0.5}$, 
and an optical-to-AGN bolometric luminosity conversion factor of 
$\rm L_{bol}=10.4\,\nu L_{\nu,4400\AA}$ (Richards et al. 2006). 
In figure 4, We plot all the z$\sim$6 quasars, as well as the 
local IR quasar sample and the average luminosities of the 
submillimeter/millimeter [(sub)mm]-observed z$\sim$2 to 5 quasar samples  
(Omont et al. 2003; Hao et al. 2008). The average FIR 
luminosities of the z$\sim$2 to 5 quasars are derived with the 
mean flux densities at 250 GHz or 350 GHz (Omont et al. 2001, 2003; 
Carilli et al. 2001; Priddey et al. 2003a), adopting the same 
assumptions of dust temperature and emissivity index used 
for the z$\sim$6 quasars (See Section 3.2). 
We fit a line to the average luminosities of all these 
(sub)mm-observed quasar samples at z$\sim$2 to 6 using the 
Ordinary Least Square method (Isobe et al. 1990). This yields 
an relationship of: \\
\begin{equation}
\rm log\,\left (\frac{L_{FIR}}{L_{\odot}}\right )=(0.62\pm0.09)log\left (\frac{L_{bol}}{L_{\odot}}\right )+(3.90\pm1.29)\\
\end{equation}
between the average FIR and AGN emission in these high-redshift optically-selected quasars, and the 
non-linear slope of 0.62 might suggest contributions from both host galaxy 
star formation and AGN power to the FIR emission (Beelen et al. 2004; Netzer et al. 2007;
Hao et al. 2005, 2008; Wang et al. 2008b; Serjeant \& Hatziminaoglou 2009; Lutz et al. 2010)

The MAMBO-250 GHz observations at sub-mJy sensitivity have detected the 
most FIR luminous objects among the z$\sim$6 quasars (Bertoldi et al. 2003a, 
Petric et al. 2003; Wang et al. 2007, 2008, this work). Most of these objects follow the shallower luminosity 
correlation trend of the local IR quasars and the (sub)mm-detected quasars at z$\sim$2 to 5. 
In particular, the five MAMBO-detected z$\sim$6 quasars 
with $\rm m_{1450}\geq20.2$ (Willott et al. 2007;
Wang et al. 2008b; this work) overlap with the high luminosity
end of the IR quasars. This may suggest a starburst-dominant 
FIR emission in these most FIR luminous quasars at z$\sim$6, 
similar to that found in the local IR quasars. 
A fit to all the (sub)mm detections in the high-z  
quasar sample and the local IR quasars gives a relationship of 
\begin{equation}
\rm log\,\left (\frac{L_{FIR}}{L_{\odot}}\right )=(0.45\pm0.03)log\left (\frac{L_{bol}}{L_{\odot}}\right )+(6.62\pm0.39)\\
\end{equation} 
 
\section{Discussion: star formation in the MAMBO-detected faint z$\sim$6 quasars}

Strong FIR dust continuum emission has been 
detected in the host galaxies of $\gtrsim$1/4 of the faint z$\sim$6  
quasars ($\rm m_{1450}\geq20.2$, Willott et al. 2007; Wang et al. 2008b, 2010, this work).
The dust continuum and molecular CO detections reveal a few 
$\rm \times10^{8}\,M_{\odot}$ of FIR-emitting warm dust 
and $\rm 10^{10}\,M_{\odot}$ of molecular gas in the host galaxies of these faint 
quasars, which are comparable to the typical values found with the 
millimeter-detected optically bright SDSS z$\sim$6 quasars.
The dust formation process associated with 
the evolution of massive stars on time scales $\rm\lesssim1Gyr$ has been discussed widely 
since the discovery of $\rm 4\times10^8\,M_{\odot}$ of dust 
in one of the most luminous/massive z$\sim$6 quasars, J1148+5251 
(Beelen et al. 2006; Morgan \& Edmunds 2003; Maiolino et al. 2004; Dwek et al. 2007). 
The detections of comparble amounts of dust in these faint z$\sim$6 
quasars may suggest similar fast dust formation mechanism in their host galaxies.  
Their FIR luminosities 
estimated from the millimeter measurements are 3 to 
$\rm5\times10^{12}\,L_{\odot}$, falling on  
the FIR-to-AGN luminosity trend defined by the local IR quasars 
and the FIR luminous quasars at z$\sim$2 to 5 (Section 4.2). The lower limit of their average FIR-to-radio 
emission ratio of $\rm q>1.6$ derived with the 250 GHz and 1.4 GHz measurements is 
within the typical range of $\rm 1.6<q<3.0$ in star forming galaxies (Section 4.1).

All these results suggest significant star formation in the 
millimeter-detected optically faint quasars at the 
highest redshift. Conservatively assuming 
that 50\% of the FIR emission is powered by host galaxy star
formation, the average star formation rate (SFR) of the five 
millimeter-detected $\rm m_{1450}\geq20.2$ quasars is about 
560 $\rm M_{\odot}\,yr^{-1}$ (Kennicutt 1998). Three of them have been detected in 
molecular CO (6-5) line emission (Wang et al. 2010; this work) 
and the ratios between FIR and CO luminosities are 
220 to 370 $\rm L_{\odot}/K\,km\,s^{-1}\,pc^{2}$, which is within the
typical range found with the submillimeter galaxies and CO-detected quasars at lower redshifts 
(Solomon \& Vanden Bout 2005; Greve et al. 2005). 
This result suggests a high star formation efficiency in these CO-detected faint z$\sim$6 
quasars, similar to that 
of the extreme starburst systems (Solomon \& Vanden Bout 2005).
Additionally, if the MAMBO-detected faint z$\sim$6 quasars are indeed less massive SMBH systems 
accreting at their Eddington limit, the higher 
$\rm \left\langle L_{FIR}\right\rangle/\left\langle L_{1450}\right\rangle$ 
found with these object (see Table 4) may suggest a higher ratio 
between SFR and SMBH accretion rate in these systems compared 
to the MAMBO detections among the optically bright z$\sim$6 quasars. 
This may indicate that the stellar bulge grows faster than the 
central SMBH in these faint/less massive quasar-host 
systems at z$\sim$6. 

\section{Conclusions}

We present observations of millimeter and radio continuum and CO(6-5) 
line emission from the host galaxies of quasars at z$\sim$6. 
The new observations complete our MAMBO dust continuum survey of 
all the UV/optically faint z$\sim$6 quasars ($\rm m_{1450}\geq20.2$) discovered 
from SDSS. Combining with previous data, we calculate the average FIR and 
radio emission for a sample of 18 z$\sim$6 quasars with $\rm m_{1450}\geq20.2$. 
The mean FIR-to-AGN UV luminosity ratio of this faint quasar sample is 
about two times higher than that of the bright z$\sim$6 quasars ($\rm m_{1450}<20.2$). 
A fit to the average FIR and AGN bolometric luminosities of both the UV/optically
faint and bright z$\sim$6 quasars, and the average luminosities
of samples of submillimeter/millimeter-observed quasars at
z$\sim$2 to 5, yields a relationship of $\rm L_{FIR}\sim {L_{bol}}^{0.62}$. 
The millimeter observations have detected the most FIR luminous 
objects among these faint z$\sim$6 quasars. The 250 GHz dust continuum detections in five of the 18 
faint quasars and CO (6-5) line detections in three of them (see Table 1 and 2)  
reveal dust masses of a few $\rm 10^{8}\,M_{\odot}$ and molecular gas 
masses of $\rm 10^{10}\,M_{\odot}$ in the quasar host galaxies, which 
are comparable to the dust and gas masses found in the bright z$\sim$6 
quasars (Beelen et al. 2006; Bertoldi et al. 2003b; Carilli et al. 2007; 
Wang et al. 2008a, 2010). Their FIR luminosities are estimated to be 
$\rm 2.6\sim5.5\times10^{12}\,L_{\odot}$, and the FIR and AGN bolometric 
luminosities follow a shallower relationship of $\rm L_{FIR}\sim {L_{bol}}^{0.45}$, 
in agreement with the starburst-AGN systems at lower redshifts. 
All these results suggest active star formation in the host galaxies 
of the millimeter-detected and UV/optically faint quasars at z$\sim$6 
with SFRs of a few hundred $\rm M_{\odot}\,yr^{-1}$. Moreover, the 
higher average FIR-to AGN UV luminosity ratios (Table 4) found 
with these objects and the shallow luminosity relationship suggest  
higher SFR-to-AGN accretion rate ratios 
than that of the more luminous/massive z$\sim$6 quasars. 
Further high-resolution imaging of
the dust and molecular gas components (e.g., with ALMA) in the
quasar host galaxies will be critical
to constrain the host galaxy dynamical mass, black hole-spheroidal
host mass ratio, gas and star formation rate surface density of
these objects. 

\acknowledgments 
This work is based on observations carried out with the Max Planck Millimeter
Bolometer Array (MAMBO) on the IRAM 30m telescope, the Plateau 
de Bure Interferometer, and the Very Large Array (NRAO). IRAM is supported 
by INSU/CNRS (France), MPG (Germany) and IGN (Spain). The National 
Radio Astronomy Observatory (NRAO) is a facility of the National
Science Foundation operated under cooperative agreement by Associated 
Universities, Inc. We acknowledge support from the Max-Planck Society
and the Alexander von Humboldt Foundation through the Max-Planck-Forschungspreis 
2005. Dominik A. Riechers acknowledges support from from NASA through Hubble
Fellowship grant HST-HF-51235.01 awarded by the Space Telescope Science 
Institute, which is operated by the Association of Universities for 
Research in Astronomy, Inc., for NASA, under contract NAS 5-26555. 
M. A. Strauss thanks the support of NSF grant AST-0707266.

{\it Facilities:} \facility{IRAM: 30m (MAMBO)}, \facility{VLA}, \facility{IRAM: Interferometer}

\begin{table}
{\scriptsize \caption{Summary of the new observations}
\begin{tabular}{lccccccc}
\hline \noalign{\smallskip}
\hline \noalign{\smallskip}
Name & redshift & $\rm m_{1450}$ & $\rm S_{250GHz}$ & $\rm S_{1.4GHz}$ 
& $\rm I\Delta v_{CO(6-5)}$ & FWHM & $\rm S_{con}$\\
     &   &        & mJy  &   $\rm \mu Jy$  &  $\rm Jy\,km\,s^{-1}$ & $\rm km\,s^{-1}$ & mJy \\
(1) & (2) & (3) & (4) & (5) & (6) & (7) & (8)  \\
\noalign{\smallskip} \hline \noalign{\smallskip}
SDSS J012958.51$-$003539.7 &  5.7794$\pm$0.0008 & 22.28 & {\bf 2.37$\pm$0.49} & -- & 0.37$\pm$0.07 & 283$\pm$87 & 0.14$\pm$0.04 \\
SDSS J020332.39+001229.3 &  5.72  & 20.97 & {\bf 1.85$\pm$0.46} & {\bf 195$\pm$22$^{a}$} & $<$0.32 & -- & $<$0.12 \\
SDSS J023930.24$-$004505.4 &  5.82  & 22.28 & 1.29$\pm$0.54 & -- & -- & -- & -- \\
ULAS J131911.29+095051.4 &  6.1321$\pm$0.0012 & 19.65 & {\bf 4.20$\pm$0.65} & {\bf 64$\pm$17} & 0.43$\pm$0.09 & 537$\pm$123 & 0.31$\pm$0.08 \\
CFHQS J150941.78$-$174926.8 & 6.12 & 19.82 & 1.00$\pm$0.46$^{b}$ & 23$\pm$18 &  -- & -- & -- \\
SDSS J205321.77+004706.8 &  5.92 & 21.20 & 0.09$\pm$0.63 & -- & -- & -- & -- \\
SDSS J214755.40+010755.0 &  5.81 & 21.65 & $-$0.36$\pm$0.61 & $-$28$\pm$18 & -- & -- & -- \\
SDSS J230735.40+003149.0 &  5.87 & 21.73 & $-$0.56$\pm$0.53 & $-$21$\pm$17 & -- & -- & -- \\
SDSS J235651.58+002333.3 &  6.00 & 21.77 & 0.21$\pm$0.49 & -- & -- & -- & -- \\
\noalign{\smallskip} \hline
\end{tabular}\\
}
{\scriptsize Note -- Column (1), name; Column (2), redshift. CO redshifts are 
presented for the two sources detected by the PdBI, and we list the redshifts 
measured with the UV quasar emission lines from the discovery papers for the 
other six sources (Jiang et al. 2008, 2009); Column (3), 
rest-frame 1450 $\rm \AA$ magnitudes from the discovery papers (Mortlock 
et al. 2008, Jiang et al. 2008, 2009). Column (4), MAMBO measurements of 
the 250 GHz dust continuum; Column (5), VLA measurements of the 1.4 GHz radio continuum; 
Column (6) and (7), line flux and FWHM of the CO (6-5) line emission; 
Column (8), measurement of the dust continuum at the CO (6-5) line frequency. 
The 250 GHz and 1.4 GHz detections are marked as boldface.\\
$^{a}$Wang et al. (2008b); $^{b}$Willott et al. 2007.
}
\end{table}
\begin{table}
{\scriptsize \caption{Previous observations of the $m_{1450}\geq20.2$ z$\sim$6 quasars}
\begin{tabular}{lcccccccc}
\hline \noalign{\smallskip}
\hline \noalign{\smallskip}
Name&redshift&$\rm m_{1450}$&$\rm S_{250GHz}$&$\rm S_{1.4GHz}$&$\rm {L_{FIR}}^{b}$&$\rm M_{dust}$&$\rm M_{gas}$&Reference \\
     & & & mJy & $\mu$Jy & $\rm 10^{12}\,L_{\odot}$ & $\rm 10^{8}\,M_{\odot}$ & $\rm 10^{10}\,M_{\odot}$ & \\
(1) & (2) & (3) & (4) & (5) & (6) & (7) & (8) & (9) \\
\noalign{\smallskip} \hline \noalign{\smallskip}
SDSS J000552.34$-$000655.8&5.85 &20.83&0.36$\pm$0.48&40$\pm$130&$<$3.4&--&--&1,2\\
CFHQS J003311.40$-$012524.9&6.13&21.78$^{a}$&{\bf 1.13$\pm$0.36}&$-$27$\pm$19&2.6$\pm$0.8&1.5&--&2,3\\
SDSS J030331.40$-$001912.9&6.08&21.28&0.23$\pm$0.51&$-$85$\pm$62&$<$3.5&--&--&2\\
SDSS J035349.76+010405.4&6.05&20.22&1.20$\pm$0.46&17$\pm$19&$<$3.2&--&--&2\\
NDWFS J142516.30+325409.0&5.89&20.62$^{a}$&{\bf 2.27$\pm$0.51}&20$\pm$20&5.4$\pm$1.2&3.0&2.0&2,4\\
FIRST J142738.59+331242.0&6.12&20.33$^{a}$&0.39$\pm$0.66&{\bf 1730$\pm$131}&$<$4.6&--&--&2,5\\
SDSS J163033.90+401209.6&6.05&20.64&0.80$\pm$0.60&14$\pm$15&$<$4.2&--&--&2,6\\
CFHQS J164121.64+375520.5&6.05&21.30$^{a}$&0.08$\pm$0.46&$-$30$\pm$32&$<$3.2&--&--&2,3\\
SDSS J205406.42$-$000514.8&6.04&20.60&{\bf 2.38$\pm$0.53}&17$\pm$23&5.5$\pm$1.2&3.1&1.2&2,4\\
SDSS J231546.36$-$002357.5&6.12&21.34&0.28$\pm$0.60&31$\pm$16&$<$4.1&--&--&2\\
CFHQS J232908.28$-$030158.8&6.42&21.65$^{a}$&0.01$\pm$0.50&14$\pm$22&$<$3.4&--&--&2\\
\noalign{\smallskip} \hline
\end{tabular}\\
Note --Reference:(1) Wang et al. 2007; (2) Wang et al. 2008b; (3) Willott et al. 2007; 
(4) Wang et al. 2010; (5) McGreer et al. 2006; (6) Bertoldi et al. 2003a.\\
$^{a}$The values are calculated from the absolute 1450 $\AA$ magnitudes in 
the discovery papers (Cool et al. 2006; McGreer et al. 2006; Willott et al. 2007).
The 250 GHz and 1.4 GHz detections are marked as boldface.
$^{b}$The upper limits of $\rm L_{FIR}$ are derived with the 3$\sigma$ upper 
limits of the 250 GHz flux densities.
}
\end{table}
\begin{table}
{\scriptsize \caption{Derived parameters for the new observations}
\begin{tabular}{lcccccc}
\hline \noalign{\smallskip}
\hline \noalign{\smallskip}
Name & $\rm L_{FIR}$ & $\rm M_{dust}$ & $\rm L_{1.4GHz}$ & $\rm L'_{CO(6-5)}$ & $\rm L'_{CO(1-0)}$ & $\rm M_{gas}$ \\
     & $\rm 10^{12}\,L_{\odot}$ & $\rm 10^{8}\,M_{\odot}$& $\rm L_{\odot}\,Hz^{-1}$  & $\rm 10^{10}\,K\,km\,s^{-1}\,pc^2$ & 
$\rm 10^{10}\,K\,km\,s^{-1}\,pc^2$ & $\rm 10^{10}\,M_{\odot}$\\
(1) & (2) & (3) & (4) & (5) & (6) & (7) \\
\noalign{\smallskip} \hline \noalign{\smallskip}
J0129-0035 & 5.2$\pm$0.9 & 2.9 & --              & 1.2$\pm$0.2 & 1.5$\pm$0.3 & 1.2    \\
J0203+0012 & 4.4$\pm$1.1 & 2.5 & 0.118$\pm$0.013 & $<$1.0      & $<$1.3      & $<$1.0 \\ 
J0239-0045 & $<$3.9  & --  & --              & --          & --          & --     \\ 
J1319+0950 &10.0$\pm$1.3 & 5.7 & 0.045$\pm$0.012 & 1.5$\pm$0.3 & 1.9$\pm$0.4 & 1.5    \\
J1509$-$1749 & $<$3.2 & -- & $<$0.038 & --          & --          & --     \\
J2053+0047 & $<$4.4  &-- & --              & --          & --          & --     \\
J2147+0107 & $<$4.4  &-- & $<$0.034        & --          & --          & --     \\
J2307+0031 & $<$3.8  &-- & $<$0.032        & --          & --          & --     \\
J2356+0023 & $<$2.8  & -- & --              & --          & --          & --     \\
\noalign{\smallskip} \hline
\end{tabular}
}
\end{table}

\begin{table}
{\scriptsize \caption{Average FIR and radio emission of the z$\sim$6 quasars}
\hskip -0.3in
\begin{tabular}{lccccccccc}
\hline \noalign{\smallskip}
\hline \noalign{\smallskip}
 Group & Number$^{a}$ & $\rm \left\langle L_{1450}\right\rangle ^{b}$ & $\rm \left\langle f_{250GHz}\right\rangle$
& $\rm <L_{FIR}>$ & $\rm \frac{\left\langle L_{FIR}\right\rangle}{\left\langle L_{1450}\right\rangle}$
& Number$^{c}$ & $\rm \left\langle f_{1.4GHz}\right\rangle$ & $\rm <L_{1.4GHz}>^{d}$ & q \\
    &     & $\rm 10^{12}\,L_{\odot}$ & mJy & $\rm 10^{12}\,L_{\odot}$ & &  & $\rm \mu Jy$ & $\rm L_{\odot}\,Hz^{-1}$ & \\
(1) & (2) & (3) & (4) & (5) & (6) & (7) & (8) & (9) & (10) \\
\noalign{\smallskip} \hline \noalign{\smallskip}
\multicolumn{10}{c}{Optically faint z$\sim$6 quasars with $\rm m_{1450}\ge20.2$} \\
\noalign{\smallskip} \hline \noalign{\smallskip}
All objects & 18 & 4.3 & 0.85$\pm$0.12  & 2.0$\pm$0.3 & 0.46  & 14 & 19$\pm$6 & 0.012$\pm$0.003 & 1.63$\pm$0.15\\
(radio quiet)& 16 & 4.0 & 0.79$\pm$0.13 & 1.9$\pm$0.3 & 0.46  & 12 & 2$\pm$6 & $<$0.013 & $>$1.57 \\
\noalign{\smallskip}
250 GHz detections & 5 & 4.3 & 1.85$\pm$0.20 & 4.3$\pm$0.5 & 1.01  & 4 & 45$\pm$10 & 0.030$\pm$0.07 & 1.59$\pm$0.11 \\
(radio quiet) & 4 & 4.2 & 1.86$\pm$0.23 & 4.3$\pm$0.5 &1.03    & 3 & 1$\pm$12      & $<$0.024      &  $>$1.68 \\
\noalign{\smallskip}
250 GHz non-detections & 13 &4.3 & 0.33$\pm$0.15 & $<$1.8 & $<$0.42  & 10 & 7$\pm$7 & $<$0.019 & -- \\
(radio quiet) & 12 & 3.9  & 0.32$\pm$0.15  & $<$1.8 &$<$0.46  & 9 & 2$\pm$7 & $<$0.015 & -- \\
\noalign{\smallskip} \hline \noalign{\smallskip}
\multicolumn{10}{c}{Optically bright z$\sim$6 quasars with $\rm m_{1450}<20.2$} \\
\noalign{\smallskip} \hline \noalign{\smallskip}
All objects &  22 &17.6  & 1.64$\pm$0.13 & 3.8$\pm$0.3 &0.22  & 22 & 50$\pm$3 & 0.033$\pm$0.002 & 1.49$\pm$0.04 \\
(radio quiet)& 21 & 16.9 & 1.68$\pm$0.13 & 3.9$\pm$0.3 &0.23  & 21 & 38$\pm$3 & 0.026$\pm$0.002 &1.61$\pm$0.05  \\
\noalign{\smallskip}
250 GHz detections & 8 & 19.1 &2.71$\pm$0.18 & 6.3$\pm$0.4 &0.33 & 8   & 43$\pm$4 & 0.029$\pm$0.003 & 1.77$\pm$0.05 \\
\noalign{\smallskip}
250 GHz non-detections & 14 & 16.8 & 0.55$\pm$0.18 & 1.3$\pm$0.4 & 0.08 & 14 & 59$\pm$5 & 0.040$\pm$0.003 & 0.94$\pm$0.15 \\
(radio quiet) &  13&15.5 & 0.58$\pm$0.18 & 1.4$\pm$0.4 & 0.09 & 13 & 32$\pm$5 & 0.021$\pm$0.003 & 1.23$\pm$0.15 \\
\noalign{\smallskip} \hline
\end{tabular}\\
}
{\scriptsize 
Note--We calculate the average flux densities using 
$\rm \left\langle f_{\nu}\right\rangle={\sum w_{i}f_{\nu,i}}/{\sum w_{i}}$, 
$\rm w_{i}=1/\sigma^{2}_{\nu,i}$, and 
$\rm \left\langle \sigma_{\nu}\right\rangle=(\sum w_{i})^{-0.5}$ 
(Omont et al. 2003), where $\rm f_{\nu,i}$ and $\rm \sigma_{\nu,i}$ are the measurement and 1$\sigma$ rms
for each object at 250 GHz or 1.4 GHz. 
For results with $\rm \left\langle f_{\nu}\right\rangle$ less than 3$\rm \left\langle \sigma_{\nu}\right\rangle$, 
we take the stacking average value plus 
3$\rm \left\langle\sigma_{\nu}\right\rangle$ as the upper 
limits of the average flux densities and calculate the corresponding upper limits for the luminosities.\\
$^{a}$ Number of sources observed at 250 GHz.
$^{b}$ Average luminosity at rest-frame 1450 $\rm \AA$, derived from $\rm m_{1450}$.
$^{c}$ Number of sources observed at 1.4 GHz.
$^{d}$ A radio spectral index of -0.75 (Condon 1992) is adopted here to calculate the rest frame 1.4 GHz lunimosity.
}
\end{table}
\begin{figure}[h]
\includegraphics[height=2.6in]{fig1a.eps}
\vskip -2.8in
\hskip 3.5in
\includegraphics[height=2.5in,angle=-90]{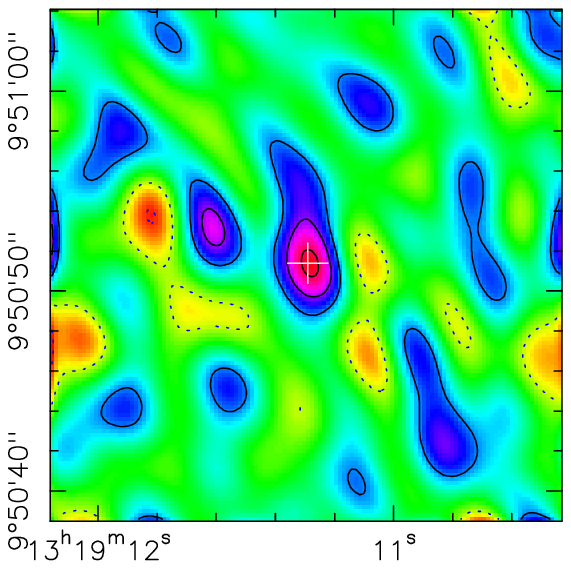}
\caption{Left -- VLA 1.4 GHz continuum image of the z=6.132 quasar J1319+0950. 
The beam size is $\rm 1.61''\times1.45''$,which is plotted on the bottom 
left, and the contour levels are (-3, -2, 2, 3, 4)$\times$15 $\rm \mu Jy\,beam^{-1}$. 
Right -- PdBI 97 GHz continuum image of the quasar averaged over the line-free 
channels. The synthesized beam size (FWHM) is 3.5$''$, and the contour levels 
are (-2, -1, 1, 2, 3)$\times$0.08 $\rm mJy\,beam^{-1}$. The cross in both 
plots denotes the optical position of the quasar (Mortlock et al. 2008).}
\end{figure}

\begin{figure}[h]
\includegraphics[height=2.0in,width=6.0in]{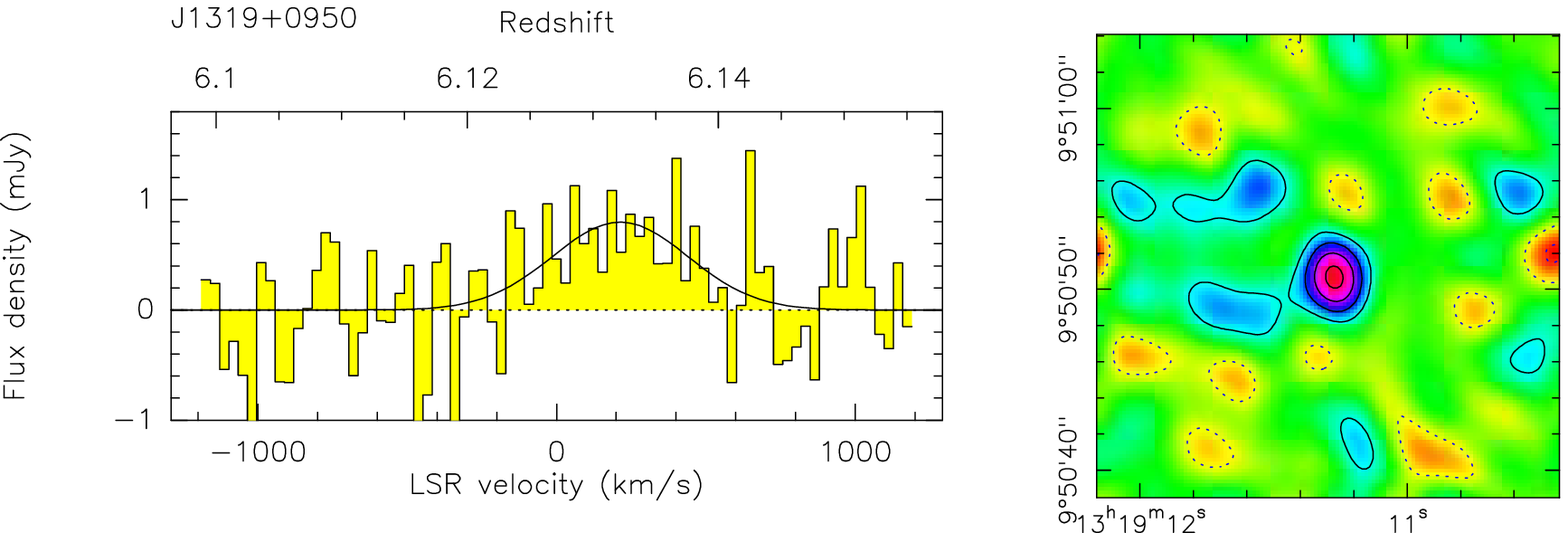}
\includegraphics[height=2.0in,width=6.0in]{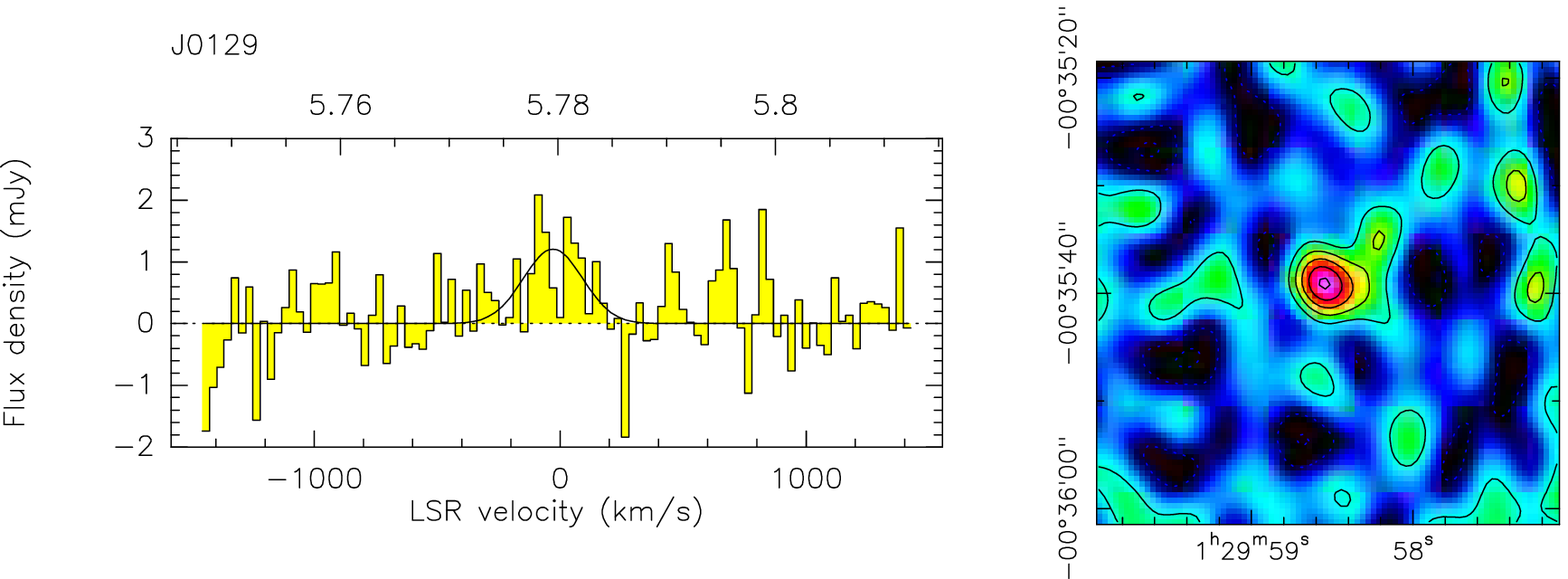}
\vskip -3.0in
\includegraphics[height=6.0in,width=5.0in,angle=-90]{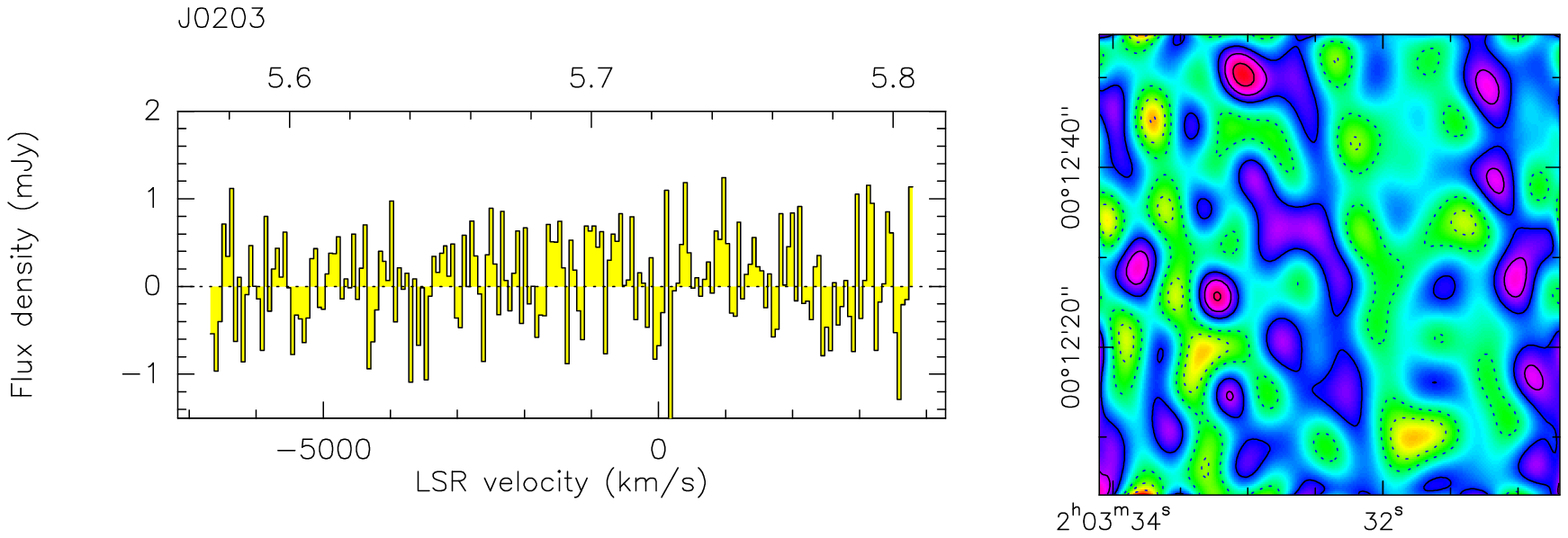}
\caption{The CO line spectra (left) and velocity-integrated
images (right) of the three z$\sim$6 quasars observed with the PdBI. 
The continuum emission underlying the CO spectra is 
removed for the two detections, J1319+0950 and J0129$-$0035. 
The synthesized beam size (FWHM) of the CO velocity-integrated image  
of J1319+0950 is $\sim$3.5$''$, and $\rm \sim5''$ for J0129$-$0035 and J0203+0012.
The top abscissae of the CO spectra give the redshift range of
the spectral windows, and zero velocity in each panel corresponds to the
optical/near-infrared redshifts of z=6.127, 5.780, 5.720 from the 
discovery papers (Mortlock et al. 2008; Jiang et al. 2009). 
The solid lines in the spectra show Gaussian fits to the line emission. 
The CO peak positions are consistent with the optical quasars for 
the two detections. 
}
\end{figure}

\begin{figure}[h]
\plotone{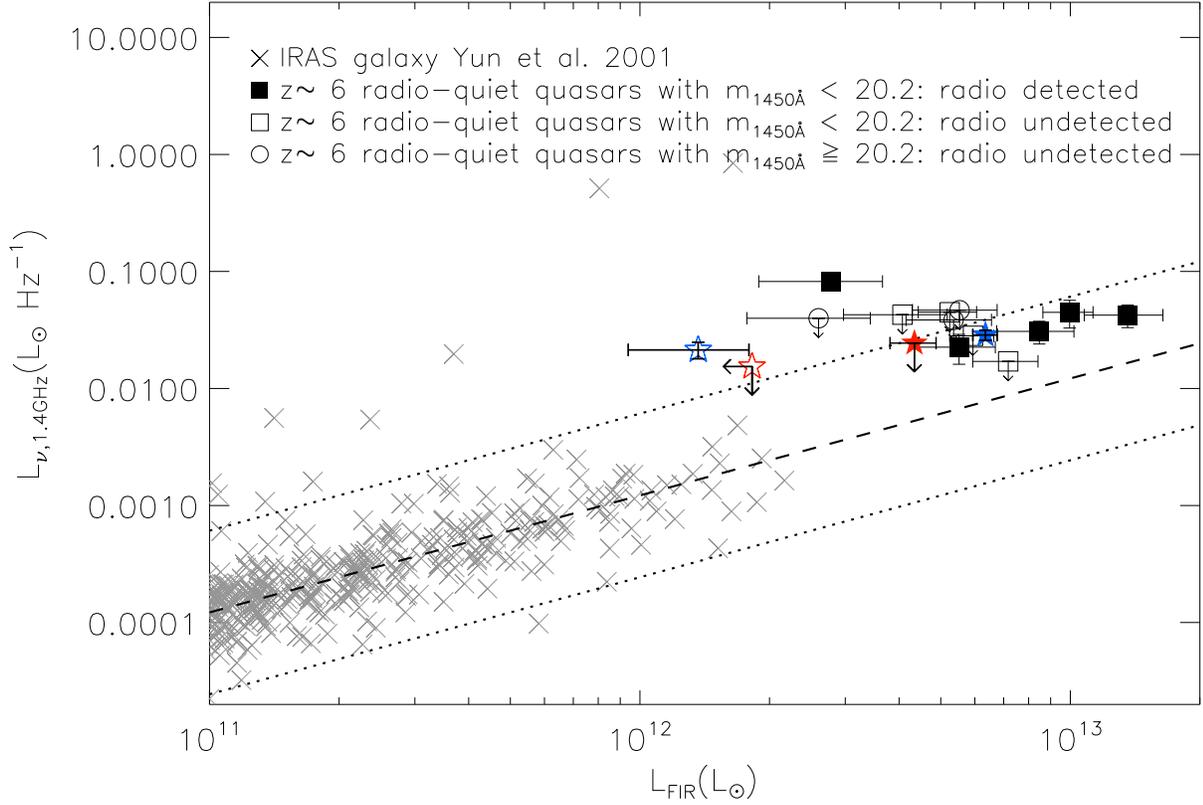}
\caption{The FIR versus radio luminosity plot of the MAMBO-detected radio-quiet z$\sim$6 
quasars that have radio 1.4 GHz continuum observations. 
The filled blue and red 
stars are the average values for the MAMBO-detected optically bright ($\rm m_{1450}<20.2$)
and faint ($\rm m_{1450}\geq20.2$) z$\sim$6 quasars, respectively. The open 
stars represent the stacking average or upper limits of the 
non-detections in the two samples. 
Only the radio-quiet objects in the z$\sim$6 quasar samples are plotted here.
The crosses are a sample of IRAS star forming 
galaxies from Yun et al. (2001). The dashed line 
shows the average radio-FIR correlation in star forming galaxies 
($\rm q=log\left(L_{FIR}/3.75\times10^{12}\,L_{\odot}\right)-
log\left(L_{1.4GHz}/L_{\odot}\cdot Hz^{-1}\right)=2.34$, Yun et al. 2001). The dotted lines
represents q values five times higher and lower, which inlcudes 98\% of the IRAS galaxies.
}
\end{figure}

\begin{figure}[h]
\plotone{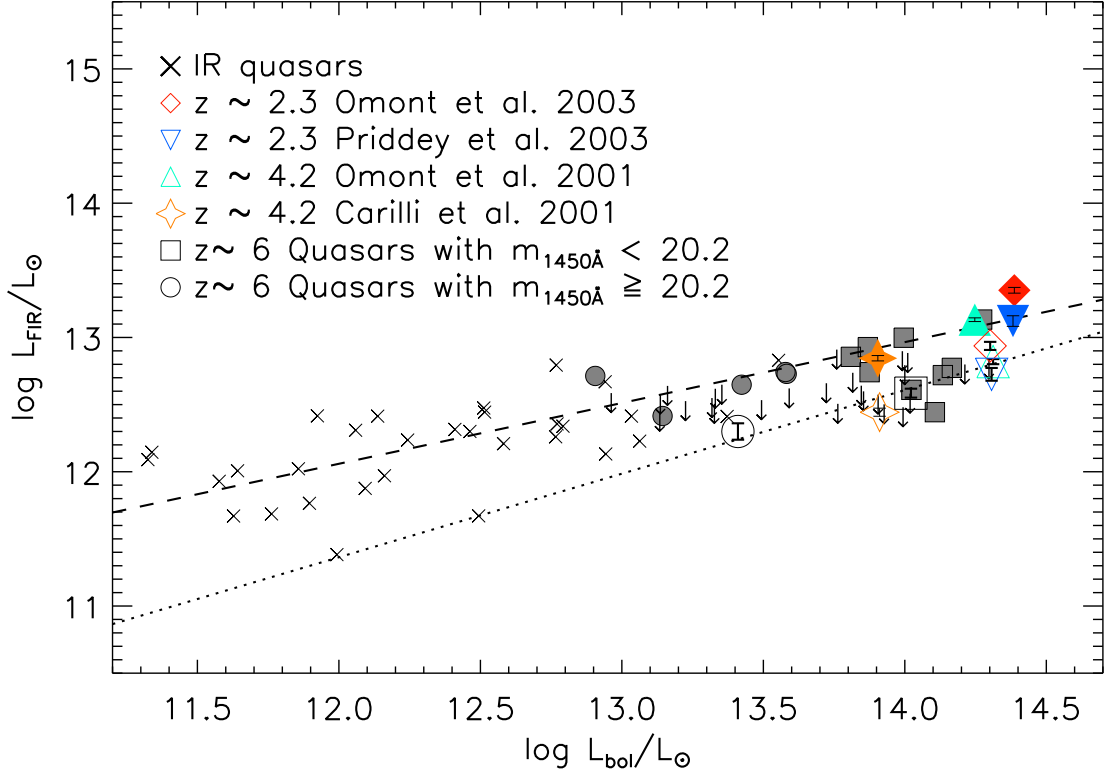}
\caption{The FIR and AGN bolometric luminosity correlations
of the z$\sim$6 quasars, together with samples of  
local optically selected IR luminous quasars hosted by ULIRGs
(Hao et al. 2005; Zheng et al. 2002) and the average luminosities of 
the (sub)mm-observed quasars at redshift z$\sim$2 to 5 (Omont et al. 2001, 2003; 
Priddey et al. 2003a; Carilli et al. 2001). 
The filled squares and circles are the MAMBO detections in the 
bright ($\rm m_{1450}<20.2$) and faint ($\rm m_{1450}\geq20.2$) z$\sim$6 quasar 
samples, respectively, and the arrows indicate 3$\sigma$ upper limits
of the non-detections (see Section 3.2). The z=6.0 quasar, SDSS 
J130608.26+035626.3, was not detected at 250 GHz (Wang et al. 2007), 
but was detected by SCUBA at 350 GHz (Priddey et al. 2003b). We estimate 
the FIR luminosity of this object with the SCUBA flux density and plot 
it as filled square. The filled triangles and diamonds are the 
mean FIR luminosities for the (sub)mm detections in each sample, 
derived by averaging flux densities at 250 GHz or 350 GHz. 
The open symbols represent the mean luminosities averaged with 
both the millimeter detections and non-detections in each sample. 
The dotted line is a fit to the average luminosities (open symbols) of all the high-z 
samples, i.e. $\rm log(L_{FIR})=0.62log(L_{bol})+3.9$, and the dashed line 
shows the fit to the submillimeter or millimeter detected 
sources in all high-z samples and the local ULIRGs, $\rm log(L_{FIR})=0.45log(L_{bol})+6.6$.}
\end{figure}

\end{document}